# Disentangling the Complex Electronic Structure of an Adsorbed Nanographene: Cycloarene C108

Jose Martinez-Castro*[1,6], Rustem Bolat[1,2,3], Qitang Fan[4], Simon Werner[4], Hadi H. Arefi[1,2], Taner Esat[1,2], Jörg Sundermeyer[4], Christian Wagner[1,2], J. Michael Gottfried[4], Ruslan Temirov[1,2,5], Markus Ternes*[1,2,6], F. Stefan Tautz[1,2,3]

[1]Peter Grünberg Institut (PGI-3), Forschungszentrum Jülich, 52425 Jülich, Germany.

[2]Jülich Aachen Research Alliance (JARA), Fundamentals of Future Information Technology, 52425 Jülich, Germany.

[3]Institut für Experimentalphysik IV A, RWTH Aachen, 52074 Aachen, Germany.

[4]Fachbereich Chemie, Philipps-Universität Marburg, 35032 Marburg, Germany.

[5]II. Physikalisches Institut, Universität zu Köln, 50937 Köln, Germany

[6]Institut für Experimentalphysik II B, RWTH Aachen, 52074 Aachen, Germany.

* Corresponding authors. Email: j.martinez@fz-juelich.de, m.ternes@fz-juelich.de

ABSTRACT

We combine low-temperature scanning tunneling imaging and spectroscopy with CO functionalized tips and algorithmic data analysis to investigate the electronic structure of the molecular cycloarene C108 (graphene nanoring) adsorbed on a Au(111) surface. We demonstrate that CO functionalized tips enhance the visibility of molecular resonances, both in differential conductance spectra and in real-space topographic images. Comparing our experimental data with ab-initio density functional theory reveals a remarkably precise agreement of the molecular orbitals and enables us to disentangle close-lying molecular states only separated by 50 meV at an energy of 2 eV below the Fermi level. We propose this combination of techniques as a promising new route for a precise electronic characterization of complex molecules and other physical properties which have electronic resonances in the tip-sample junction.

**INTRODUCTION**

Molecular adsorbates are characterized by electronic resonances that can be spectroscopically detected[1]. Scanning tunneling microscopy (STM) with its sub-molecular imaging resolution and its ability to probe the local density of states (LDOS) by recording the differential conductance d$I$/d$V$ spectra yields spatially resolved images of molecular states[2,3]. Since scanning tunneling spectroscopy (STS) can access occupied as well as unoccupied molecular states, it has become a valuable tool to unveil the electronic structure of molecular adsorbates[4,5].

So far, the spatial mapping of molecular resonances has been conducted mostly in the following way: The tunneling bias of the junction is set to the energy value at which the resonance occurs and, while the molecule is scanned by the STM tip, the d$I$/d$V$ signal is recorded. Such scanning procedure can be performed in two distinct modes: constant-height or constant-current[6]. The obtained d$I$/d$V$ map is commonly interpreted as a two-dimensional visualization of the molecular state. This approach has been successfully employed to molecular systems on which electronic resonances are well-separated from each other and in many cases decoupled from the metallic substrate by ultrathin insulators[3,7,8], revealing a clear correspondence between calculated molecular orbitals and the measured electronic resonances. However, for the cases when the molecular resonance spectrum is denser and the detected electronic resonances have contributions from different molecular states[9], the interpretation of the d$I$/d$V$ data becomes a difficult problem which has in general not been solved yet.

In this work, we use a graphene nanoring, the cycloarene C108 adsorbed on Au(111), as a model system. Cycloarenes are considered as ideal test cases for studying electron delocalization in extended aromatic systems[10]. The C108 possesses a rich energy spectrum in which several molecular states are only separated in energy by a few meV[11], making it particularly interesting for disentangling the contributions from the different molecular orbitals in the STM images.

To access, map and have a complete understanding of the C108 electronic structure, we use an approach that aims to overcome the limitation of common d$I$/d$V$ imaging. For that, we combine

CO functionalized tips, shown to provide sub-molecular resolution particularly on $sp^2$-hybridized carbon nanostructures[12–14], with feature-detection scanning tunneling spectroscopy (FD-STS)[15], a statistical analysis tool that allows the unbiased detection of electronic resonances. We find that this combination enables us to classify, resolve and separate orbitals lying close in energy, providing experimental information on the electronic structure of a cycloarene in unprecedented detail.

**RESULTS**

**Synthesis of Cycloarene C108 on Au(111).** We synthesize the cycloarene C108 on a Au(111) surface via Ullmann coupling followed by cyclodehydrogenation of a bromoarene precursor containing 9 phenyl moieties (5'''-([1,1'-biphenyl]-2-yl)-5'',5''''-dibromo-1,1':2',1'':3''',1''':3'''',1'''':3''''',1''''':2'''''',1''''''-septiphenyl) (Fig. 1a-c). The first step, on-surface Ullmann coupling, leads to the cyclodimerization of the precursor into a nonplanar macrocycle, which subsequently forms C108. This new on-surface cyclodimerization reaction is more efficient than the on-surface cyclotrimerization used in previous work for the first synthesis of C108[11]. Details of the C108 precursor synthesis can be found in the supporting information.

**STM and STS Characterization.** We image C108 at a base temperature of 10 K using a tungsten tip functionalized with a single CO molecule (CO tip), enabling sub-molecular resolution of the geometrical structure (Fig. 1d). Using the same tip, we also perform d$I$/d$V$ spectroscopy revealing several electronic resonances in the energy range between -2eV and +2eV around the Fermi energy $E_F$ (Fig. 1e). The CO tip impacts the measured images and spectra compared to a bare metallic s-wave tip in two ways: first, at biases close to $E_F$ it allows to achieve structural sub-molecular resolution, similar to previous reports[13,16–18] and, second, it enhances significantly the intensity of the electronic resonances. We identify the four main features as positive (PIR) and negative ion resonances (NIR)[11] (orange dots in Fig. 1e). Compared to previously published results, with the increased resolution of the CO tip, we additionally identify four secondary features shifted by approximately 150 mV from the main features to higher absolute bias (green dots in Fig. 1e).

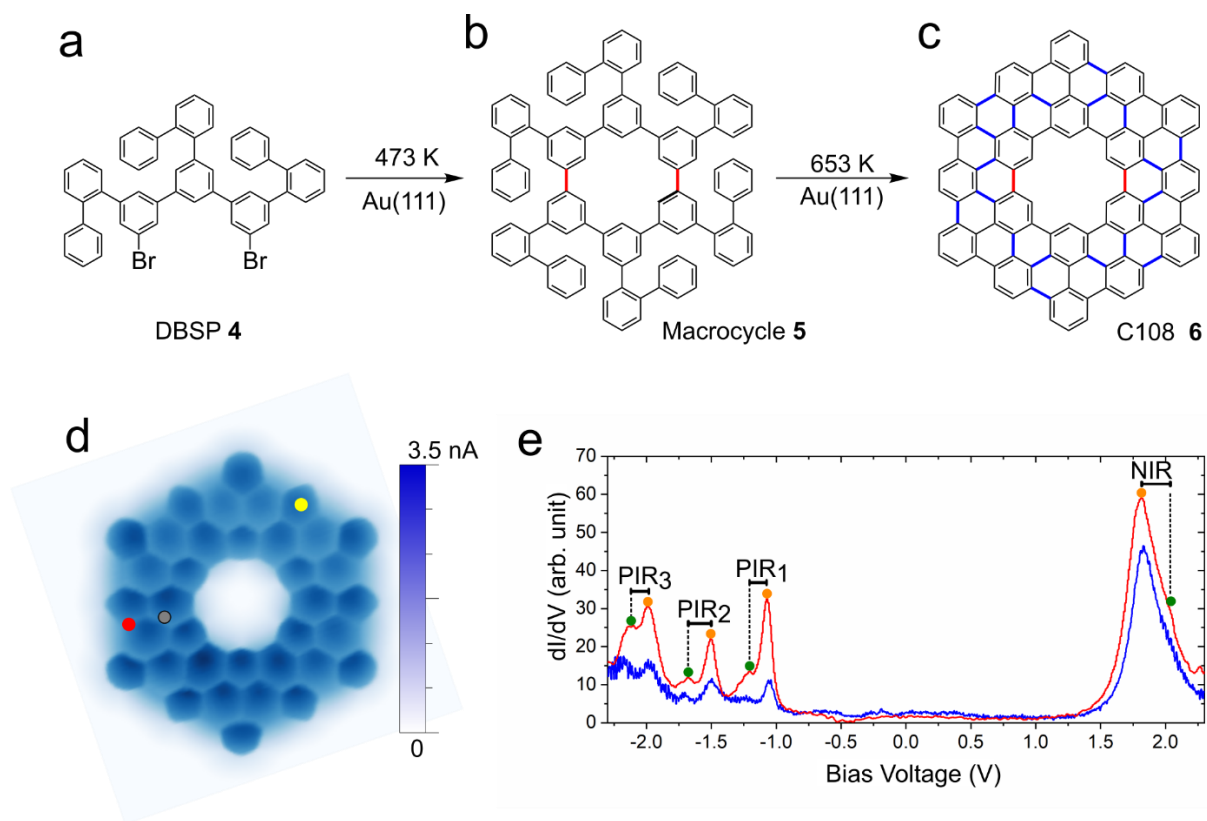

**Figure 1. Synthesis, structural and electronic properties of C108.** The precursor 5'''-([1,1'-biphenyl]-2-yl)-5'',5''''-dibromo-1,1':2',1'':3''',1''':3'''',1'''':3''''',1''''':2'''''',1''''''-septiphenyl (DBSP) **(a)** is vapor-deposited onto the surface. Subsequent annealing induces an Ullmann coupling, resulting in the nonplanar macrocycle **(b)**. Upon further annealing, the C108 macrocycle is formed by cyclodehydrogenation **(c)**. The C–C bonds formed via Ullmann and dehydrogenative coupling are shown in red and blue, respectively. **(d)** Constant-height STM image of the C108 nanoring ($V$ = 20 mV, size 2.4 × 2.4 nm$^2$). The red and gray dots mark the locations at which the spectra in Fig. 2b were recorded. **(e)** d$I$/d$V$ spectra taken with a metallic (blue) and a CO tip (red), measured over the yellow dot marked in (d). Main features (orange dots) and their replicas (green dots) are labeled with PIR$_1$, PIR$_2$, PIR$_3$ and NIR ($V$ = 200 mV, $I$ = 80 pA, $V_{mod}$ = 20 mV).

**Density Functional Calculations.** To understand the origin of the resonances, we perform density functional theory (DFT) calculations that use the B3LYP hybrid functional to calculate the density of states (DOS). From the gas phase DFT calculation (Fig. 2a,b) and by comparing with a representative d$I$/d$V$ spectrum of C108 (red curve in Fig. 2b), we find that the highest occupied molecular orbital (HOMO, H$_0$) at -1.094 eV, along with the orbitals HOMO-1 and HOMO-2 (H$_{-1}$, H$_{-2}$), situated both at -1.100 eV, are nearly degenerate and contribute to the measured resonance PIR$_1$. The degenerate orbitals HOMO-3 and HOMO-4 (H$_{-3}$, H$_{-4}$), located both at -1.503 eV, contribute to the resonance PIR$_2$ and HOMO-5, HOMO-6 (H$_{-5}$, H$_{-6}$) at -1.964 eV and -2.019 eV contribute to the resonance PIR$_3$. On the unoccupied side ($V$ > 0), we observe experimentally a broad NIR peak spanning through the energy position of three different calculated orbitals: the lowest unoccupied molecular orbital (LUMO, $L_0$) at +1.830 eV and the degenerate LUMO+1 and LUMO+2 at +1.911 eV ($L_{+1}$, $L_{+2}$).

The significant intensity variations between the NIR and the $PIR_{1-3}$, which are not found in this extent in the calculated DOS, stem from the strongly varying tunneling probabilities for tunneling into or out of these resonances. At positive $V$, the tunneling barrier height is effectively reduced, leading to an increase of the NIR, while for increasingly negative $V$, the barrier for tunneling from these resonances into the tip is successively increased[6].

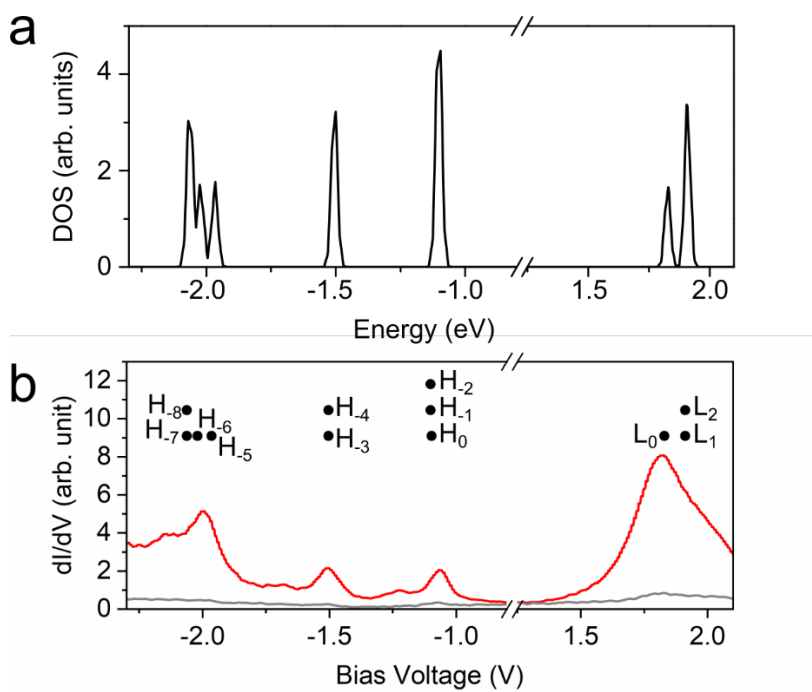

**Figure 2. DFT-calculated DOS and molecular orbital energy levels of the cycloarene C108**. **(a)** Simulated DOS of C108 in the gas phase (black curve). **(b)** Energetic positions of the gas-phase orbitals as found in the DFT calculation (black dots), overlaid with typical experimental d$I$/d$V$ spectra, red and gray, taken at the positions of the red and gray dots, respectively, in Fig. 1d. Spectroscopy parameters prior to opening the feedback: $V$ = 200 mV, $I$ = 80 pA. $V_{mod}$ = 20 mV. In both graphs, the region between -0.8 V and 1.2 V is omitted for a better visualization of the resonances.

**Molecular Orbital Imaging**. To investigate the spatial distribution of the molecular orbitals, we start by performing constant-current d$I$/d$V$ maps at the bias voltage corresponding to the center of the $PIR_2$, $PIR_1$ and NIR resonances with a CO tip at relatively low tunneling resistances $R = V/I \approx 40 - 500 M\Omega$ (Fig. 3). To compare theory and experiment, we must consider the electronic broadening due to the bias modulation $V_{mod} = 20$ mV (peak to peak) used for the lock-in detection, which limits the energy resolution. Therefore, we compute $|\nabla_r \Psi(r)|^2$ with $\nabla_r$ as the in-plane gradient and $\Psi(r)$ as the DFT calculated wavefunction and sum up those molecular orbitals which lie within the broadening range of the detection method (Fig. 3a-c). The computed p-wave images and the measured constant current d$I$/d$V$ maps match very well. The good agreement between theory (Fig. 3a-c) and experiment (Fig. 3d-f) stems from a low hybridization between the C108 and the metallic surface and the energetically well-separated resonances, so that the constant current d$I$/d$V$ maps do not capture contributions from other states, avoiding intermixing.

Remarkably, we find that the obtained constant current d$I$/d$V$ maps of PIR$_1$ and NIR (Fig. 3e, f) look very different compared to the constant height d$I$/d$V$ maps obtained by Fan et al. at presumably significantly larger tip-sample separation.[11] These differences are due to the CO tip's p-wave character at low tunneling resistances, i.e. small tip-sample separations, which is sensitive to the nodal planes of the molecular wave function[19]. The metallic s-wave character of the CO tip can be recovered by increasing the tip-sample distance and then the tunneling resistance on which the s-wave character predominates over the faster decaying p-wave character [20] (Supp. Fig. 1).

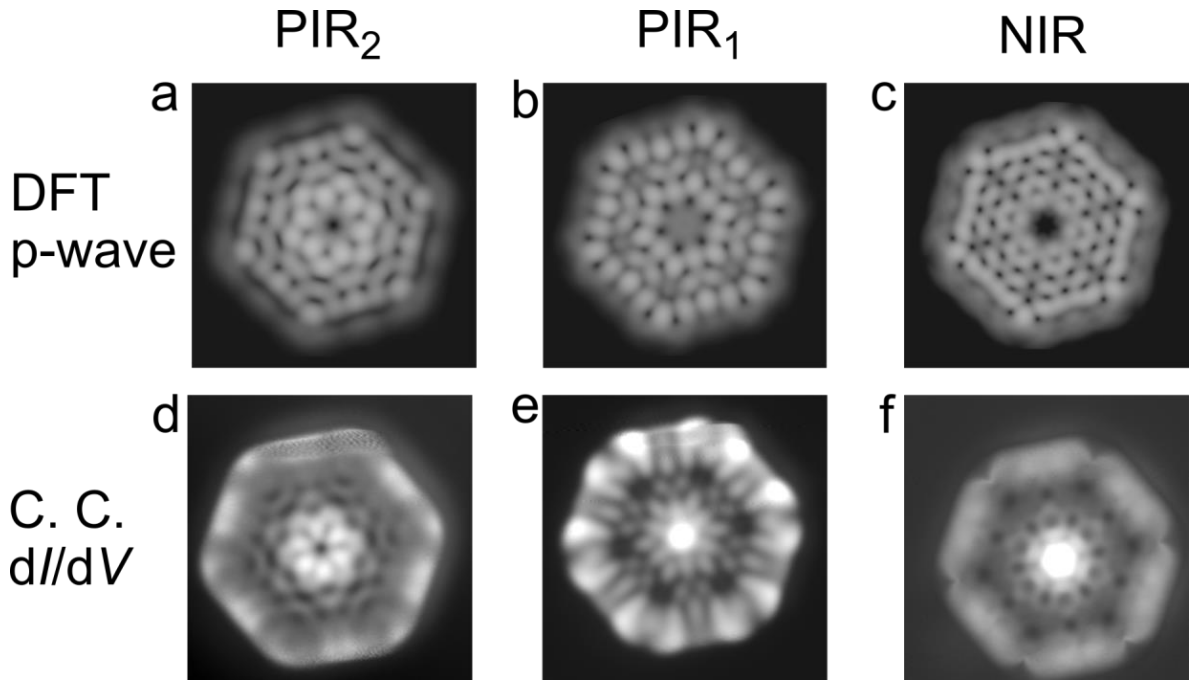

**Figure 3**. **Calculated gas-phase p-wave molecular orbitals and constant-current d$I$/d$V$ maps of C108 obtained with a CO tip.** **(a-c)** Visualization of the molecular wavefunctions as calculated by DFT using a p-wave tip at the resonance energies $V_{PIR2} = -1.52$ V, $V_{PIR1} = -1.06$ V, and $V_{NIR} = 1.8$ V. **(d-f)** Constant-current d$I$/d$V$ maps performed with CO tip at the bias voltages $V_{PIR2} = -1.47$ V, $V_{PIR1} = -1.05$ V, and $V_{NIR} = 2$ V, and current setpoints of $I_{PIR2} = 5$ nA, $I_{PIR1} = 3$ nA and $I_{NIR} = 50$ nA, $V_{mod} = 10$ mV. Image sizes are $2.5 \times 2.5$ nm$^2$.

We continue by investigating PIR$_3$ which has contributions from H$_{-5}$ and H$_{-6}$. Because in this case H$_{-5}$ and H$_{-6}$ are separated by an energy difference of 55 meV, slightly higher than the modulation voltage, we compute p-wave images of the molecular orbitals for H$_{-5}$ and H$_{-6}$ separately (Fig. 4a, b) as well as for the sum of H$_{-5}$ and H$_{-6}$ (Fig. 4c). Comparing with the constant current d$I$/d$V$ image, we find that it resembles more the sum of the two orbitals (Fig. 4d). Apparently, we see here the limit of constant current d$I$/d$V$ map which do not allow to discern between two energetically neighboring molecular orbitals.

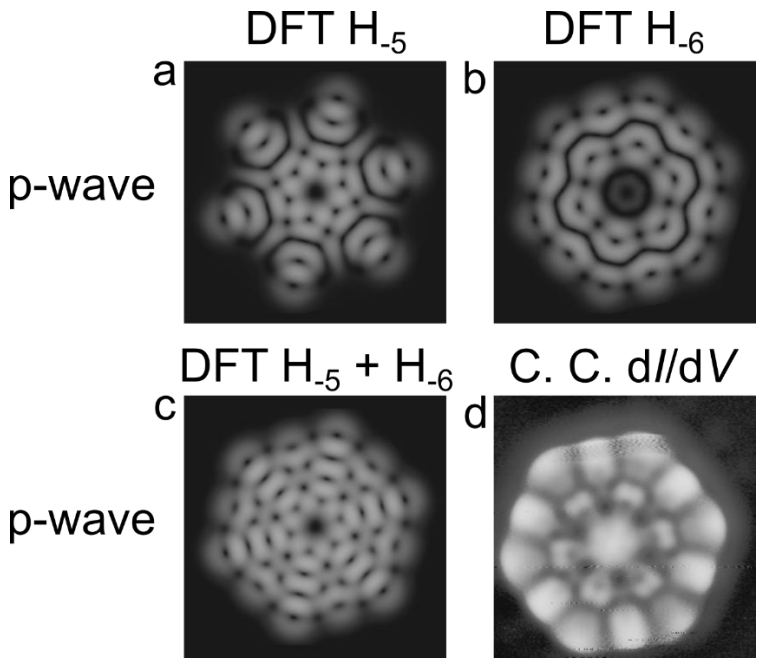

**Figure 4**. **Calculated gas-phase p-wave molecular orbitals and constant-current d$I$/d$V$ maps of C108 obtained with a CO tip of PIR$_3$.** **(a-b)** Visualization of the molecular wavefunctions as calculated by DFT using a p-wave tip of H$_{-5}$ = -1.964 eV and H$_{-6}$ = -2.019 and the sum of H$_{-5}$ and H$_{-6}$ **c)**. **d)** Constant-current d$I$/d$V$ map performed with CO tip at the bias voltage $V_{PIR3} = -1.95$ V and current setpoint of $I_{PIR3} = 40$ nA, with $V_{mod} = 10$ mV. Image sizes are $2.5 \times 2.5$ nm$^2$.

**Feature Detection Scanning Tunneling Spectroscopy Method.** Until here, we have only explored the traditional way of how electronic properties of molecules have been characterized by STM. Now, we would like to utilize a method which gives a more complete picture of the molecular electronic properties even at larger tip-sample separations where the high resolution of the CO tip is no longer predominant.

For this purpose, we measure $dI/dV$ spectra on a grid of 64 × 64 points covering an area of 2.4 × 2.4 nm$^2$ centered on an isolated C108 molecule with the CO tip. At each grid point, we regulate the tip height prior to opening the feedback loop to a tunneling current of $I = 80$ pA at a bias of $V = 200$ mV. The bias lies inside the energy gap between PIR and NIR to avoid the influence of molecular orbitals on the tip height and thus the resulting spectroscopic intensities. We then analyze the corresponding matrix of spectra using a recently developed FD-STS algorithm[15].

The feature detection enables one to find and classify spectral features in a large set of individual $dI/dV$ spectra. This method circumvents the necessity of using a model describing the physical processes and its subsequent time-consuming least-square fits. Because we are mainly interested in finding molecular states which are smeared out by hybridization with the substrate and by thermal broadening, the task is to detect broadened peaks on an unknown background (see the methods section for more details of the peak detection procedure).

To each detected peak in the raw $dI/dV$ signal, we assign a weight $W$ by finding the next lying extrema in the derivative of the signal, *i.e.* the second derivative $dI^2/d^2V$ (Figure 5a). Note that even though $W$ is not identical to the peak intensity in $dI/dV$, it is a measure commonly used, for instance, in Auger and ESR spectroscopy. By adjusting a threshold $W_c$, we disregard spurious noise peaks as well as peaks clearly originating from the substrate only (Fig. 5b).

The statistical evaluation of the features detected by FD-STS allows us to plot the (weighted) occurrence center energies $E_p$ of the detected peaks in a certain energy range across the grid in an *energy distribution histogram* (EDH) (Fig. 5c), or to explicitly map the (weighted) occurrence of spectroscopic features in a *feature distribution map* (FDM). Finally, it is possible to map the energy position of a given feature, yielding an *energy distribution map* (EDM).

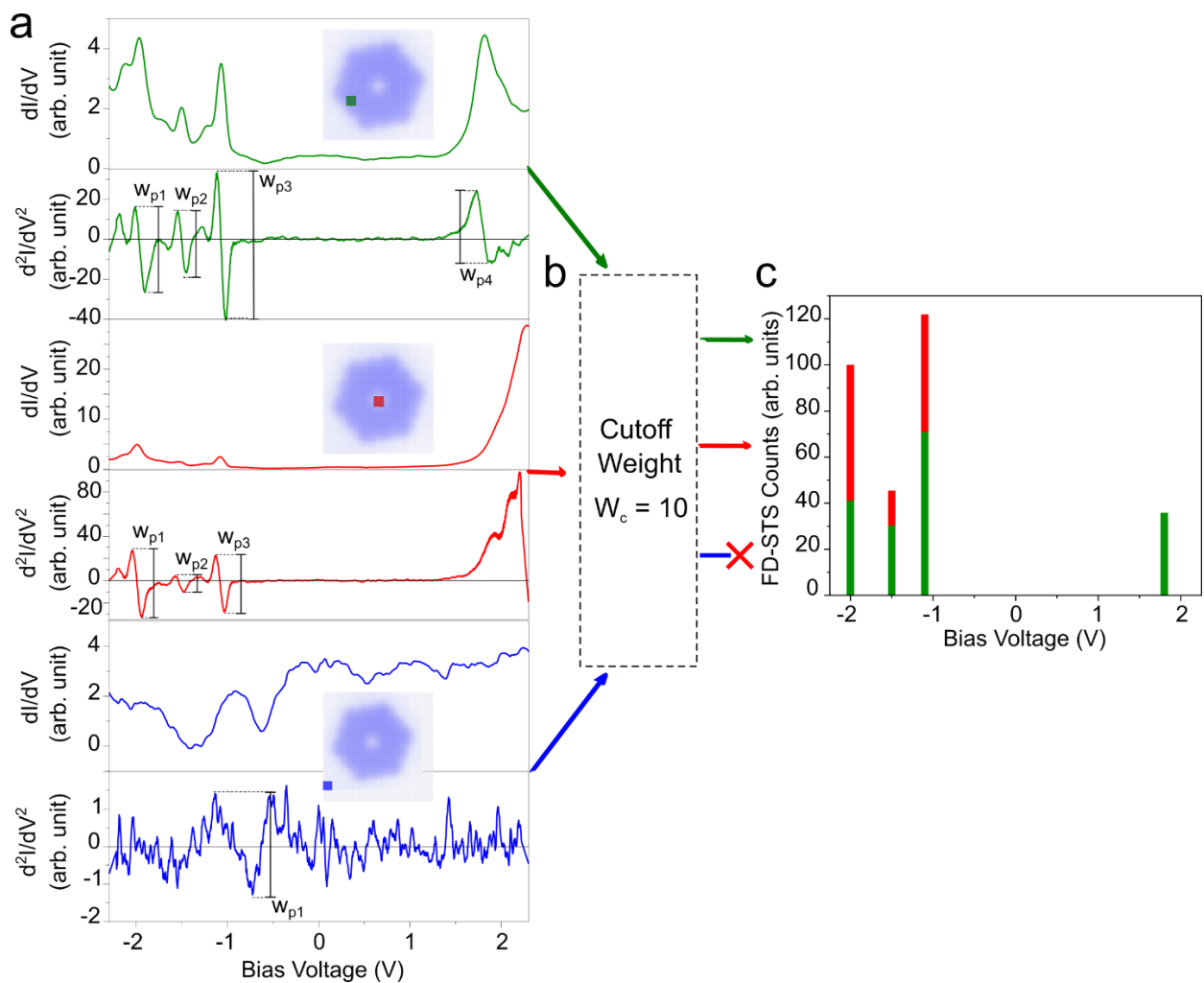

**Figure 5. Schematic representation of FD-STS algorithm.** (**a**) d$I$/d$V$ and d$^2I$/d$V^2$ spectra for different positions (edge of the molecule, center of the molecule, and outside the molecule). (**b**) Schematic representation of the criterion to either collect or discard a given feature according to its weight. (**c**) Weighted histogram for the filtered features that passed the cutoff weight.

**Energy of the molecular states.** We start our analysis by normalizing each individual spectrum to the intensity sum of the 3 PIRs that are the result of occupied molecular orbitals in the C108. This normalization compensates for the different tunneling probabilities occurring at different locations on the C108 (red and gray curves in Fig. 2b) assuming an approximately constant total electron density within the spatial extension of the molecule. By applying the FD-STS algorithm, we are able to detect the center of the peaks in each spectrum. The EDH in Fig. 6a reflects the PIR and NIR main energies and intensities (red in Fig. 6a) as well as a series of side peaks (cyan in Fig. 6a) separated by bias voltages ranging from 140 to 210 meV.

As we will show, FDMs and EDMs are particularly useful for the classification and visualization of molecular orbitals that are very close in energy and are present in different regions of the molecule. In combination with the enhanced spatial resolution of the CO tip's p-wave nature[19,21], this results in better understanding of the electronic structure than conventional d$I$/d$V$ maps.

We start by selecting the energy range of the EDH that will be used for the FDM and EDMs (yellow boxes in Fig. 6b-d). The sharpness of the selected peaks suggests that only a single resonance is contributing, despite originating from the contributions of more than one molecular orbital. The produced EDMs use a linear scale to visualize the energy distribution of the selected resonance over the previously defined energy range (Fig. 6e-g). Additionally, the EDMs confirm that the resonances are energetically monodisperse varying approximately only by the energy of the modulation broadening within the molecule. In other words, we do not observe different resonances merged to a larger resonance. Unlike in previous work[15], for the FDMs we plot the intensity at the center of each detected peak of the normalized d$I$/d$V$ (Fig. 6h-j). We notice that even by adjusting $W_c$ to zero, i.e. no filtering any peak attributed to spurious noise, or by using different window widths we obtain qualitatively the same EDH and FDM with the caveat of being noisier (Supp. Fig. 2, 3).

Next, we proceed to relate the spatial distribution of the detected features with FD-STS to molecular orbitals. In contrast to Fig. 3 and 4, we consider here an equal contribution of s and a p-wave orbitals from the CO tip due to the increased tip-sample distance at the higher tunneling setpoint $R = 2.5 G\Omega$[20]. The FDMs reproduce the main features of the calculated s + p wave images (Fig. 6). For instance, the FDM obtained for PIR$_1$ (Fig. 6i) shows lobes situated at the corners of the C108 molecule, in agreement with the corresponding DFT calculations (Fig. 6l).

We note that the molecular resonances decay in their intensity as the tip moves towards the inner part of the ring either in the constant-current d$I$/d$V$ maps or in the normalized spectra used for the FDM. This effect can be compensated by decreasing the tip-sample distance as shown in Fig. 3 and 4. However, often this is not feasible due to the fragility of adsorbed molecules. This reveals the limits of not only our approach but also to similar experiments conducted on many different molecular systems probed both with metallic s-wave [8,22] and CO tips[23], in which the molecular states are observed extending beyond the rim of the molecular structure, while the interior of the molecule tends to be electronically transparent. We observe in Fig. 6 that at larger absolute bias voltages (i.e. larger tunneling barriers) the FDMs tend to concentrate the features closer to the periphery of the molecule.

Remarkably, and contrary to the DFT calculation, the FDM as well as constant-current d$I$/d$V$ maps (Fig. 3, 4 and Fig. 6) show an increased intensity in the central hole of the C108 nanoring. In this position, the tip is placed above the Au(111) surface, which is why we attribute the detection of these electronic resonances to a closer tip-sample distance caused by the directional sharper CO tip that is able to pick up the tails of electronic contributions from the molecular orbitals. On the contrary, such intensity in the center of the C108 molecule is not observed when the same set of measurements is performed with a metallic s-wave tip (Supp. Fig. 4) due to its relative bluntness compared to a CO tip.

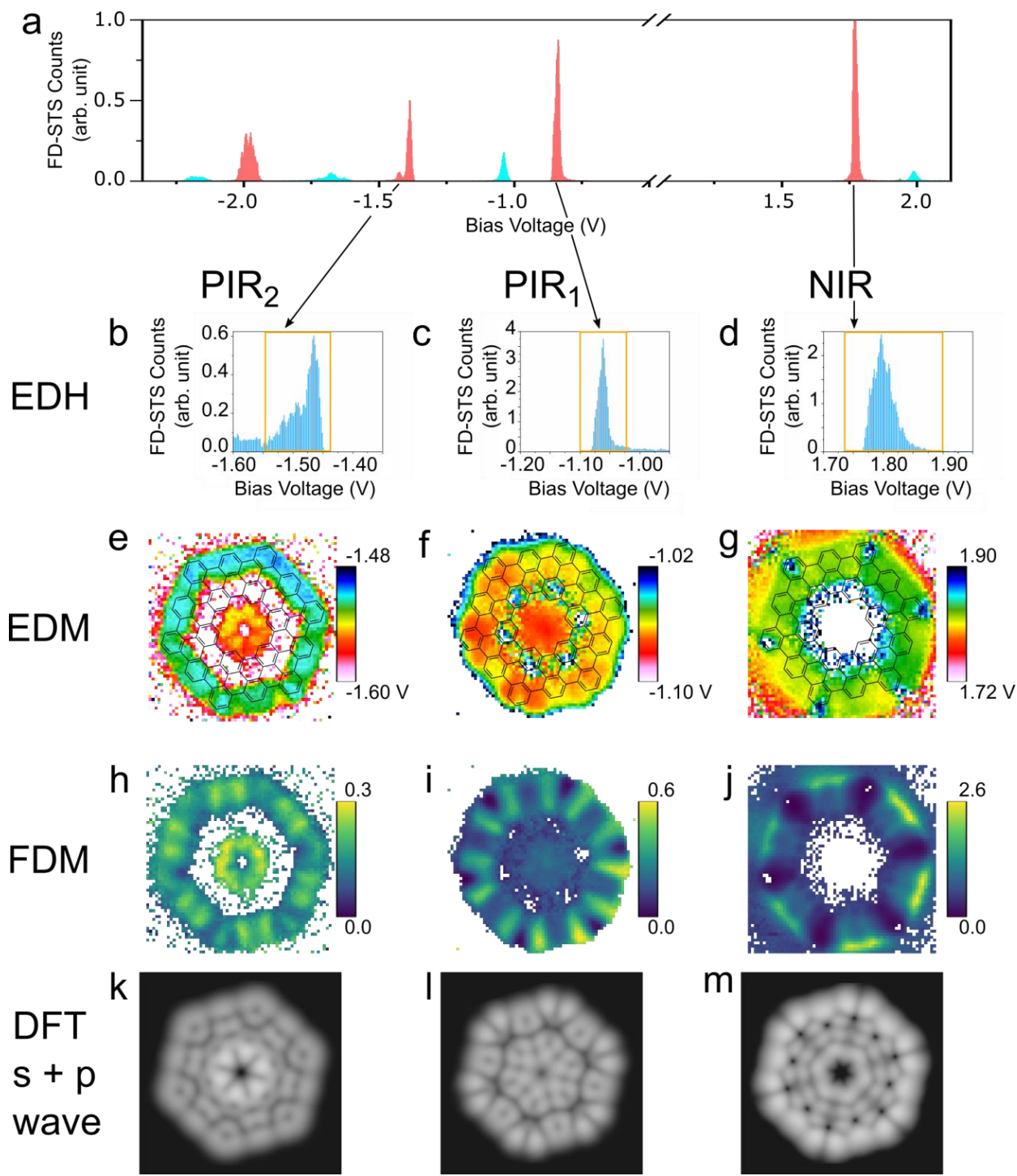

**Figure 6. FD-STS with a CO tip and calculated gas-phase molecular orbitals of C108.** **(a)** C108/Au(111) histogram of detected peaks (energy distribution histogram EDH, red) and vibronic replicas (cyan) produced by statistical analysis of a 64 × 64 grid of d*I*/d*V* spectra measured over the molecule. **(b-d)** Energy distribution histograms (EDH) of the same resonances. The yellow boxes indicate the selected energy ranges chosen for the FDMs and EDMs in e-g and h-j, respectively. **(e-g)** Energy distribution maps (EDM) of PIR$_2$, PIR$_1$ and NIR (from left to right), overlaid with the chemical structure of C108. **(h-j)**

Feature distribution maps (FDM) of PIR$_2$, PIR$_1$ and NIR (from left to right). **(k-m)** Visualization of the of the molecular wavefunctions as calculated by DFT using an sp-wave mixed tip at the resonance energies $V_{PIR2} = -1.52$ V, $V_{PIR1} = -1.06$ V, and $V_{NIR} = 1.8$ V. Image sizes are $2.4 \times 2.4$ nm$^2$.

**Vibronic replicas.** We now focus on the side peaks separated by bias voltages ranging from 140 to 210 meV from the main peaks. We attribute these peaks to vibronic replica[24–26] of the C108 electronic states. This conjecture is supported by a calculation of the vibrational spectrum based on a finite-difference approach at DFT-PBE level; the observed vibronic replica stem most likely from a family of vibrational modes found at frequencies ranging from 1100 to 1600 cm$^{-1}$ (Supp. Fig. 5), equivalent to 140 to 200 meV.

A close inspection of the vibrational replica (Supp. Fig. 6) reveals a substantial broadening of the EDH compared to the main peak, understood in terms of the symmetry dependence of the vibration assisted tunneling[27]. Additionally, by analyzing the resulting FDMs in Supp. Fig. 6, we observe that the maps reproduce qualitatively the main resonance to which it belongs, confirming that these peaks are the result of vibrational replicas.

**Disentanglement of nearby molecular orbitals.** Finally, we focus on the PIR$_3$. In contrast to the maps of Fig. 6e-g, the EDM shows a resonance that is composed of two separable regions with slightly different energies: Yellow/green areas with higher peak energy and red/orange areas with lower peak energy (Fig. 7a). A closer look at the EDH reveals that a peak which appears to be monodisperse at first glance is a composition of two separate peaks whose energetic centers are located at −2.00 and −1.95 V (Fig 7b). The strong additional maximum in between the two peaks is an artifact of the FD-STS algorithm: if none of the two resonances is dominant, the superposition of both will result in an apparent single peak close to the intermediate energy, artificially creating an extra peak in the histogram. Remarkably, our DFT calculations predict two non-degenerate molecular orbitals (H$_{-5}$ and H$_{-6}$) at approximately −2 eV with a separation of only 55 meV (Fig. 7e, f).

To separate these two orbitals, we now fit this part of the d$I$/d$V$ spectra with a sum of two Lorenzians with fixed energies at −1.95 and −2.00 eV as determined by the EDH. The resulting intensity maps (Fig. 7c, d) are in remarkable agreement with the DFT calculations (Fig. 7e,f), showing the potential of FD-STS to identify energetically overlapping molecular orbitals that can be separated to obtain a reliable image of the molecular orbitals in contrast to constant-current d$I$/d$V$ maps showing contributions of both molecular orbitals (Fig. 4d).

Since also NIR contains two orbitals, we follow the same procedure. In this case, the resonance is the contribution of $L_0$ and the degenerate $L_{+1}$ and $L_{+2}$ with an energy difference of 90 meV (Fig. 2b). We fix Lorentzian centers at an energy of 1.84 and 1.98 V, corresponding to the positions on which the central peak and the peak shoulders are found (Fig. 1b). In contrast to Fig. 7, however, we do not observe any significant difference in their intensities (Supp. Fig. 7) in agreement with the EDH (Fig. 6d), that only shows a single peak. We relate this to the lack of differences in the wavefunction intensities between the $L_0$ and the degenerate $L_{+1}$ and $L_{+2}$.

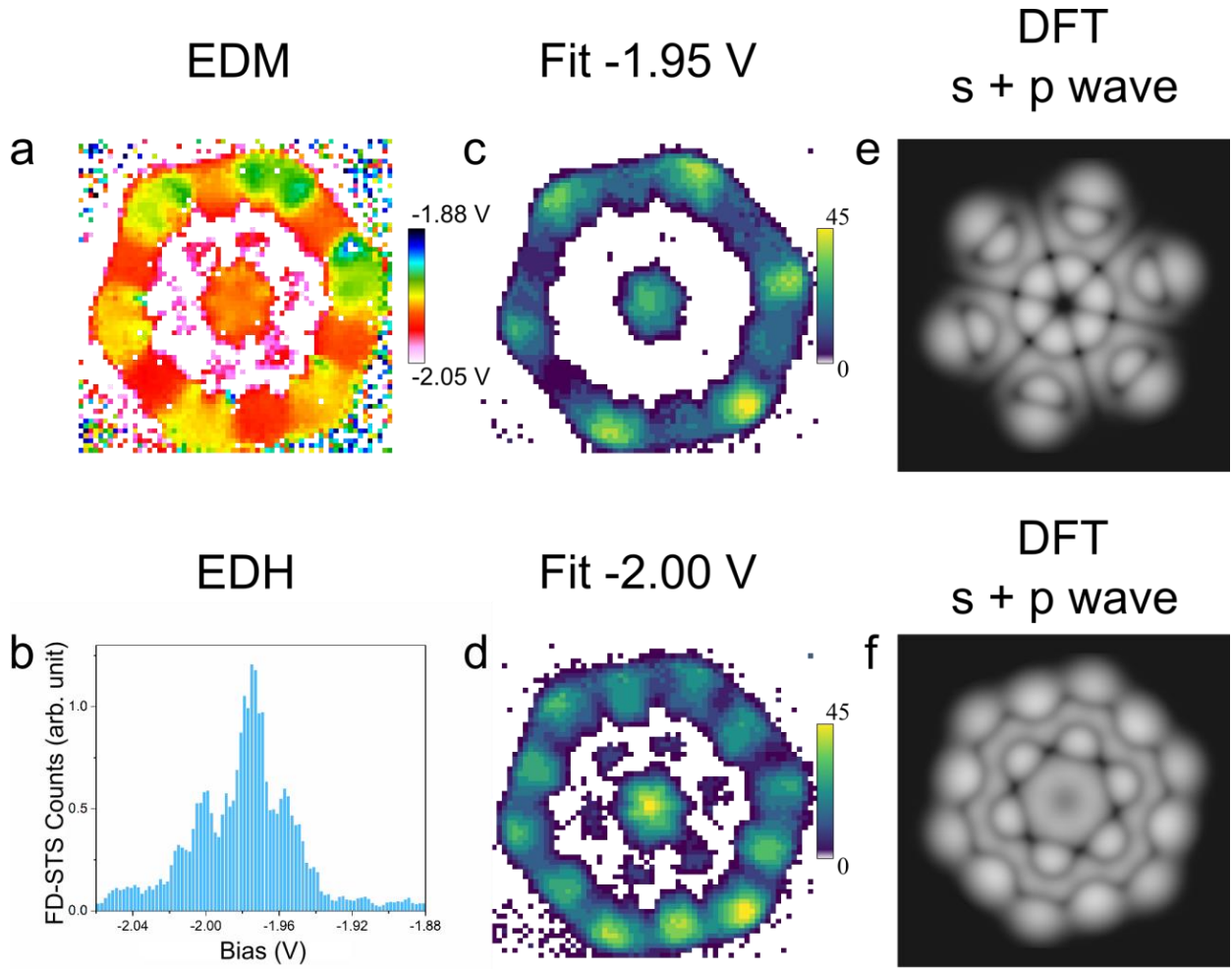

**Figure 7. Feature distribution and peak fitting maps for PIR$_3$ and calculated molecular orbitals. (a)** Feature distribution map of PIR$_3$, showing two clearly differentiated regions which is also reflected in the energy distribution histogram in **(b)**. **(c-d)** Peak intensity maps, obtained by fitting PIR$_3$ with two Lorentzians whose centers have been fixed at -1.95 V and -2 V. **(e-f)** Visualization of the molecular

wavefunctions as calculated by DFT using an sp-wave mixed tip at the resonance energies $V_{PIR3}= -1.95$ and $-2$ V. Image sizes are $2.4 \times 2.4$ nm$^2$.

## DISCUSSION

We have shown that the combination of a CO tip and FD-STS is a powerful tool to identify, classify and map the molecular states of the cycloarene C108 in good agreement with theoretical calculations. With this combination, we reach a level of detail in the understanding of the molecule's electronic properties that is neither achievable by constant-current d$I$/d$V$ maps nor by using a metallic tip and applying FD-STS alone. In particular, our method shows its potential by experimentally identifying and disentangling the close-lying molecular states at -2 V only separated by an energy difference of 50 mV. We note that this method could also be applied to systems on which different types of electronic excitations are observed, *e.g.*, superconductivity,[28] spin excitations[29] or to molecular systems such as those incorporating transition metals or rare earth atoms[30], on which theoretical calculations are limited due to their complexity. Furthermore, we expect that the methodology presented here will make the studies of the complex electronic structure of larger nanostructures more efficient. In the case of C108, the unambiguous identification of molecular resonances may help in future experiments exploring the intriguing effects of high magnetic fields such as persistent ring currents[31,32] or the Aharonov-Bohm effect[33,34].

## METHODS

**Feature Detection Algorithm.** We implement the feature detection method in the following way: The spectra containing the bare signal $S$ are in our case d$I$/d$V$ spectra obtained by modulating the swept bias voltage $V$ with a small sinusoidal modulation voltage $V_{mod} = 20$ mV (peak to peak) and recording the first derivative d$I$/d$V$ of the tunneling current $I(V)$ with the help of a lock-in detection scheme. Because these spectra are equally spaced with respect to $V$, we can smooth out their intrinsic noise by using the well-established Savitzky-Golay method, which does a polynomial fit on a sliding window with very little computational effort[35]. Since we are mainly interested in finding peaks, a second order Savitzky-Golay filter, which uses local quadratic fits of the form $S(x) = a + bx + cx^2$ on each point of all spectra is sufficient. The Savitzky-Golay filter provides us not only with the smoothed $S = a$, but also directly with the smoothed first and second derivative ($S' = b$ and $S'' = c$) of the signal $S$. To avoid filtering out meaningful signal components we use a sliding window of approximately 100 mV width, less than the full-width half-maximum of the expected peak structures. Furthermore, we check that the residuals, $R = S-a$, of the spectra do not significantly exceed the standard deviation $\sigma$ of all residuals.

To find peaks (dips) in the spectra, we search for zero crossing in $S'$ by evaluating $sign(S'(n)) - sign(S'(n+1)) < 0(0)$, with $sign(x) = x/|x|$ as the signum function and $n$ as the bin of the linearly increasing bias $V$. Note that the method can be straightforwardly extended to also detect steps as well as "shoulders": masked peaks which due to some nearby larger structures do not develop a local maximum.

**Density Functional Theory calculations.** We performed Density Functional Theory (DFT) calculations using the Fritz-Haber-Institute *ab initio* molecular simulation package (FHI-aims)[36]. The geometry of the graphene nanoring was optimized using Becke's three parameter hybrid exchange functional and the Lee Yang Parr correlation functional (B3LYP)[37–40]. Applying the Broyden–Fletcher–Goldfarb–Shanno (BFGS) algorithm, we relaxed the structure via two consecutive steps; first by expanding the Kohn-Sham wavefunction with the default numerical "light" basis sets[41], applied to all atomic species. Once finished, all basis sets were replaced with the default numerical "tight" basis sets[7] and the relaxation was resumed. We continued the relaxation until the maximum force on each atom, in either setup, was less than $10^{-3}$ eV/Å. To aid the convergence in either relaxation step, we applied a 0.05 eV broadening to all states, using a Gaussian occupation smearing scheme[42]. It was later reduced to 0.01 eV for precisely calculating the molecular orbitals and density of states. Pair-wise intramolecular dispersion effects were included using the Tkatchenko-Scheffler (TS) approach[43] and relativistic effects were accounted for via scaled zeroth-order regular approximation (ZORA)[44] To obtain a well converged electronic description of the systems, a threshold of $10^{-6}$ eV for the total energy, $10^{-4}$ eV for the sum of eigenvalues and $10^{-6}$ e/Å$^3$ for the charge density was applied during all self-consistent field (SCF) cycles.

The vibrational spectrum was computed using a finite displacement approach and assuming harmonic approximation. In order to reduce the computational costs, the semi-local exchange-correlation functional of Perdew, Burke and Ernzerhof (GGA-PBE)[45] was used to treat the electronic exchange and correlation. First the graphene nanoring was fully relaxed at DFT+PBE/tight-bs level with a lowered force threshold of $10^{-4}$ eV/Å for improved accuracy. Subsequently, the Hessian matrix was determined numerically by finite displacements of all atomic positions in six directions ± ($x$, $y$, z) by $25 \times 10^{-4}$ Å, and then diagonalized to obtain the vibrational modes. The corresponding intensities are obtained by taking the derivative of the dipole moments along these modes and broadened with Lorentzian function for visual enhancement and comparison with experimental data.

**Acknowledgements**

M.T. acknowledge funding by the Heisenberg Program (TE 833/2-1) of the Deutsche Forschungsgemeinschaft (DFG), J.M.C by the Humboldt Foundation, M.T. and J.M-C. acknowledge funding by the Priority Program (SPP 2244) of the DFG. J.M.G. and F.S.T. acknowledge funding by the DFG through SFB 1083 "Structure and Dynamics of Internal Interfaces" (223848855-SFB 1083). J.M.G. and J.S. acknowledge funding by the LOEWE Program of Excellence of the Federal State of Hesse (LOEWE Focus Group PriOSS "Principles of On-Surface Synthesis"). H.A. and C.W. acknowledge funding through the European Research Council (ERC-StG 757634 "CM3") and gratefully acknowledge the computing time granted through JARA on the supercomputer JURECA at Forschungszentrum Jülich.

**Author Contributions**

J.M.-C., R.T. and F.S.T. conceived the project. J.M.-C. and R.B. performed the experiments. J.M.-C., R.B., T.E. and M.T evaluated the data. H.A. and C.W. performed the DFT calculations. S.W. and J.S. synthesized the C108 precursor. Q.F. and J.M.G. developed the on-surface synthesis protocol. All authors discussed the results. J.M.-C and M.T. wrote the paper with significant contributions from all authors.

**Competing Interests**

The authors declare no competing financial interest.

Raw data are available at the Jülich DATA public repository.

**Supplementary Information**

Disentangling the Complex Electronic Structure of an Adsorbed Nanographene: Cycloarene C108

Authors:

Jose Martinez-Castro*[1,6], Rustem Bolat[1,2,3], Qitang Fan[4], Simon Werner[4], Hadi H. Arefi[1,2], Taner Esat[1,2], Jörg Sundermeyer[4], Christian Wagner[1,2], J. Michael Gottfried[4], Ruslan Temirov[1,2,5], Markus Ternes*[1,6], F. Stefan Tautz[1,2,3]

[1]Peter Grünberg Institut (PGI-3), Forschungszentrum Jülich, 52425 Jülich, Germany.

[2]Jülich Aachen Research Alliance (JARA), Fundamentals of Future Information Technology, 52425 Jülich, Germany.

[3]Institut für Experimentalphysik IV A, RWTH Aachen, 52074 Aachen, Germany.

[4]Fachbereich Chemie, Philipps-Universität Marburg, 35032 Marburg, Germany.

[5]II. Physikalisches Institut, Universität zu Köln, 50937 Köln, Germany

* Corresponding authors. Email: j.martinez@fz-juelich.de, m.ternes@fz-juelich.de

**Supplementary Figures.**

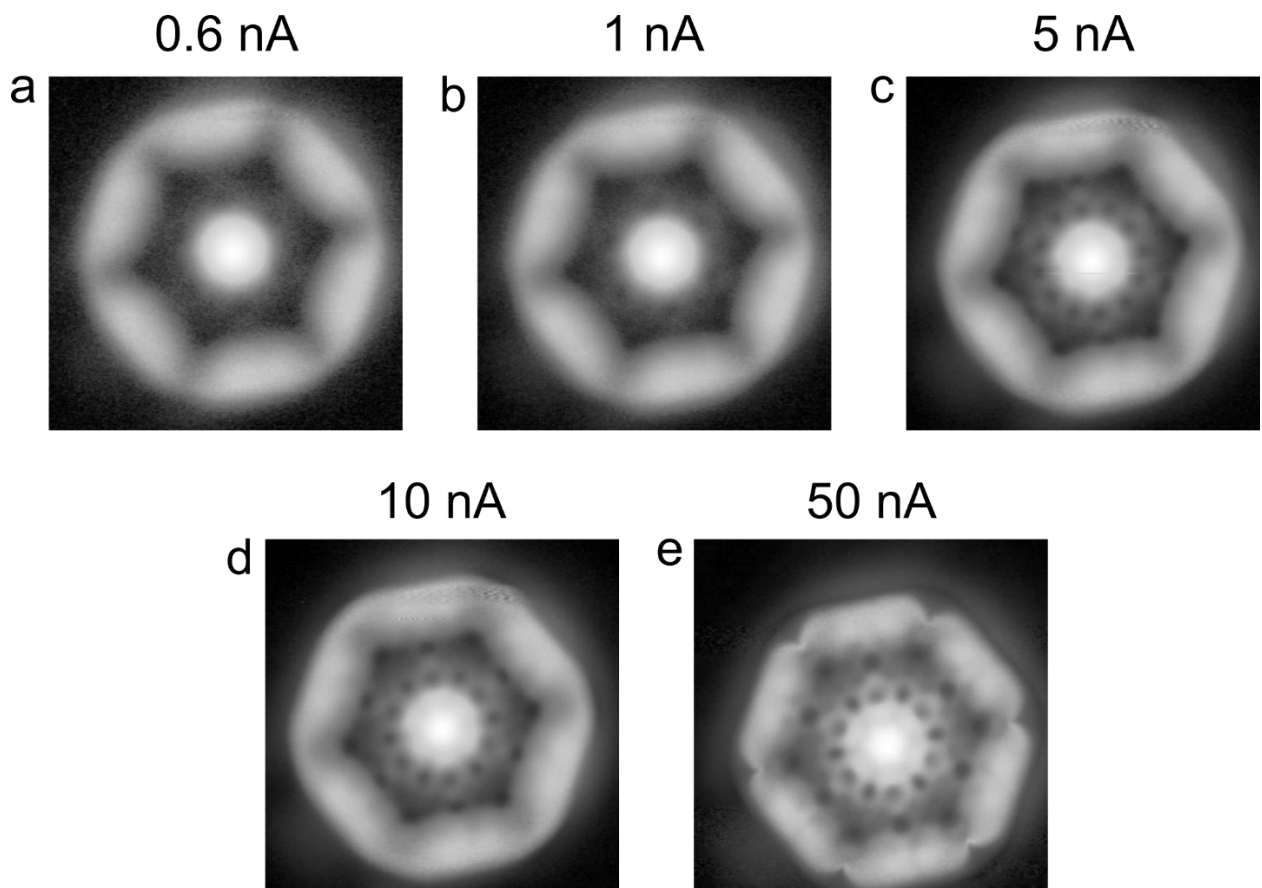

**Supplementary Figure 1**. **Variation of the negative ion resonance (NIR) image as a function of the p- and s-wave CO tip orbital predominance.** **(a-e)** Constant-current d$I$/d$V$ maps of C108 with a functionalized CO tip performed over NIR with different setpoint currents of $I$ = 0.6, 1, 5, 10 and 50 nA respectively. $V$ = 2 V, $V_{mod}$ = 20 mV. Image sizes are 2.4 × 2.4 nm$^2$.

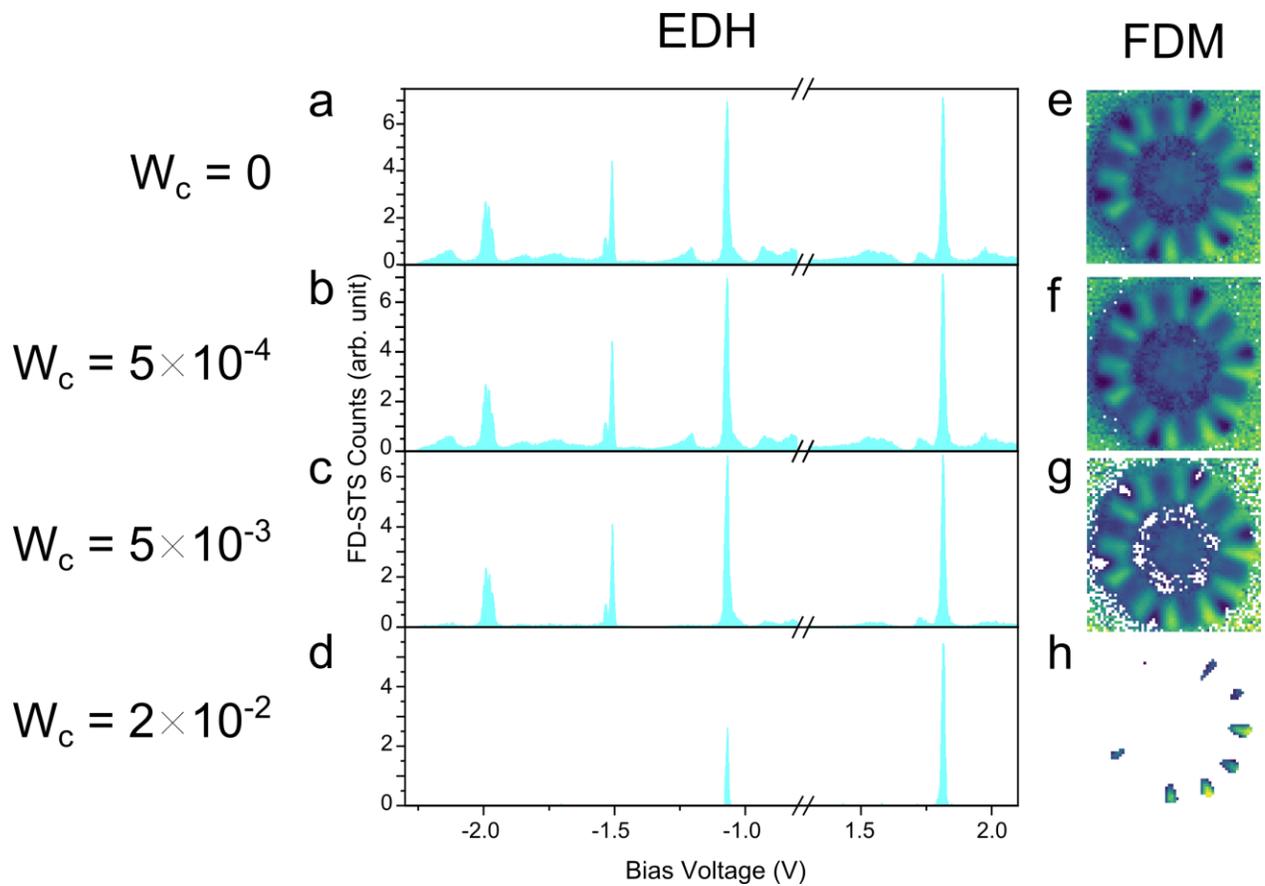

**Supplementary Figure 2**. **FD-STS of C108 as a function of the cutoff weight. (a-d)** Energy distribution histograms (EDH) of the detected peaks and feature detection maps (FDM) of $PIR_1$ **(e-h)** as a function of the cut off weight. Image sizes are $2.4 \times 2.4$ nm$^2$.

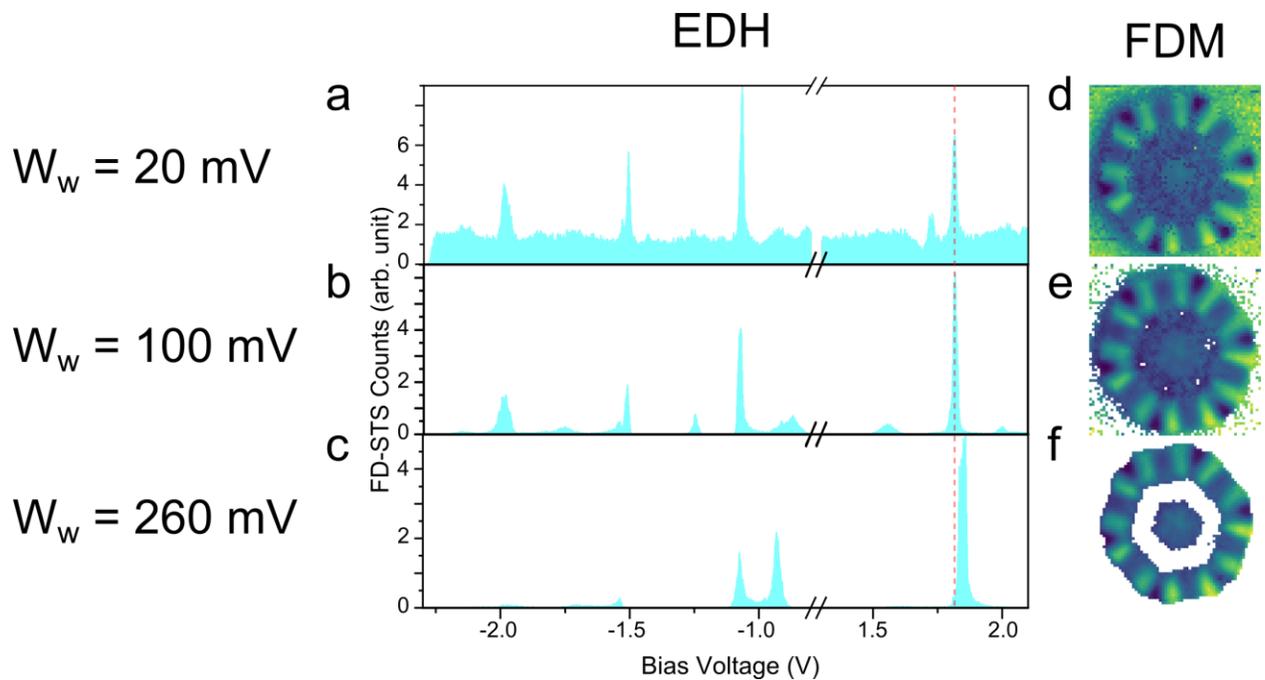

**Supplementary Figure 3**. **FD-STS of C108 as a function of the window width of the *Savitzky-Golay* method. (a-c)** Energy distribution histograms (EDH) of the detected peaks and feature detection maps (FDM) of $PIR_1$ **(d-f)** as a function of the window witdh. The red dotted line is centered on $NIR_1$ corresponding to a $W_w$ of 20 mV. Image sizes are $2.4 \times 2.4$ nm$^2$.

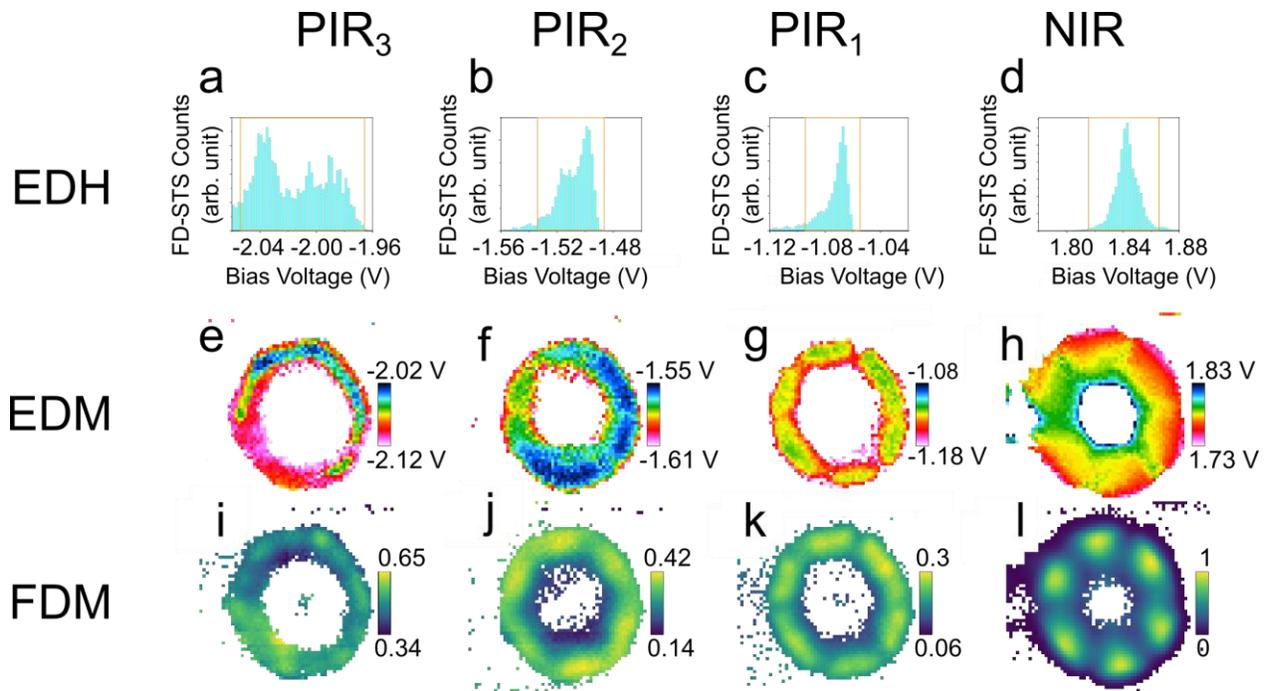

**Supplementary Figure 4. FD-STS of C108 with a metallic s-wave tip**. **(a-d)** Energy distribution histograms (EDH) of $PIR_3$, $PIR_2$, $PIR_1$ and $NIR_1$ (from left to right). The yellow boxes indicate the selected energy range chosen for the FDMs and EDMs. **(e-h)** Energy distribution maps (EDM) of $PIR_3$, $PIR_2$, $PIR_1$ and $NIR_1$ (from left to right). **(i-l)** Feature distribution maps (FDM) of $PIR_3$, $PIR_2$, $PIR_1$ and $NIR_1$ (from left to right). Image sizes are $2.4 \times 2.4$ nm$^2$.

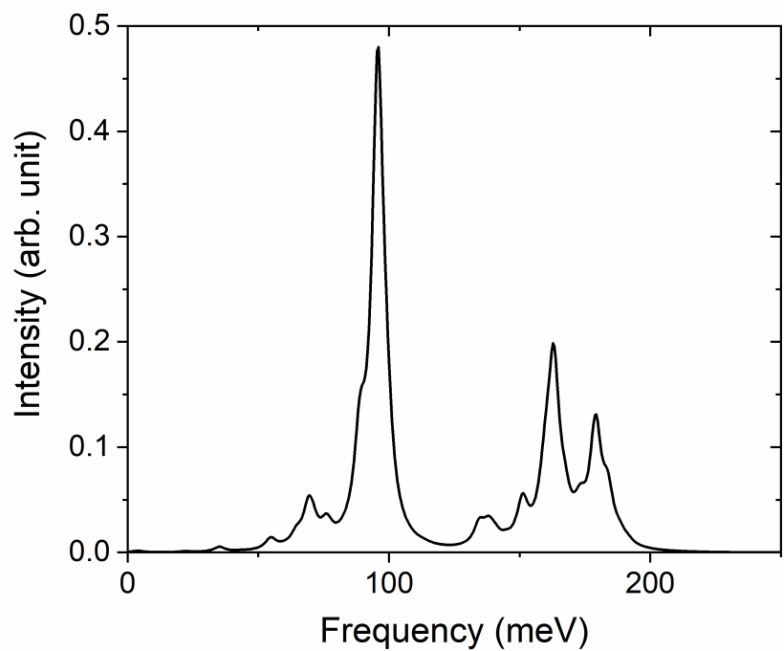

**Supplementary Figure 5**. **Calculated IR spectrum for the isolated C108 molecule.** Vibrational frequencies are obtained from diagonalization of the mass-weighted Hessian matrix, keeping only those with nonzero dipole gradients and broadened by Lorentzian functions.

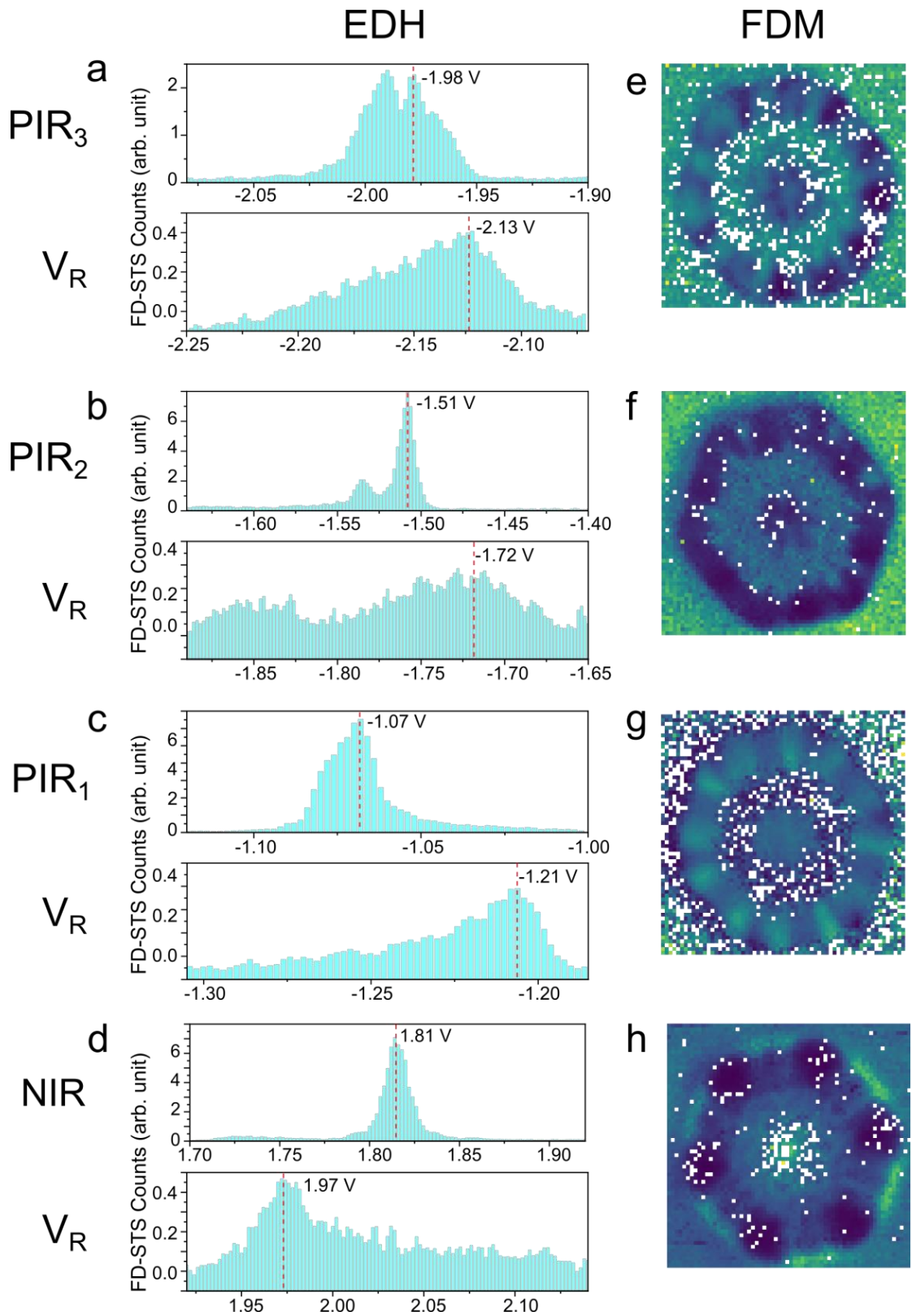

**Supplementary Figure 6**. **FD-STS of C108 vibrational replicas.** **(a-d)** Energy distribution histograms (EDH) of $PIR_3$, $PIR_2$, $PIR_1$ $NIR_1$ and their associated vibrational replica ($V_R$). The red dotted line indicates the position of the most abundant value of the EDH. **(e-h)** Feature distribution maps (FDM) of $PIR_3$, $PIR_2$, $PIR_1$, $NIR_1$ vibrational replicas ($V_R$). Image sizes are $2.4 \times 2.4$ nm$^2$. Note that the EDH associated to the main peaks are different from those showed in Fig. 6b-d and Fig. 7b because the cutoff weight $W_c$ was decreased in order to maximize FDM signal coming from the vibronic replicas and thus, contributing to the corresponding EDHs.

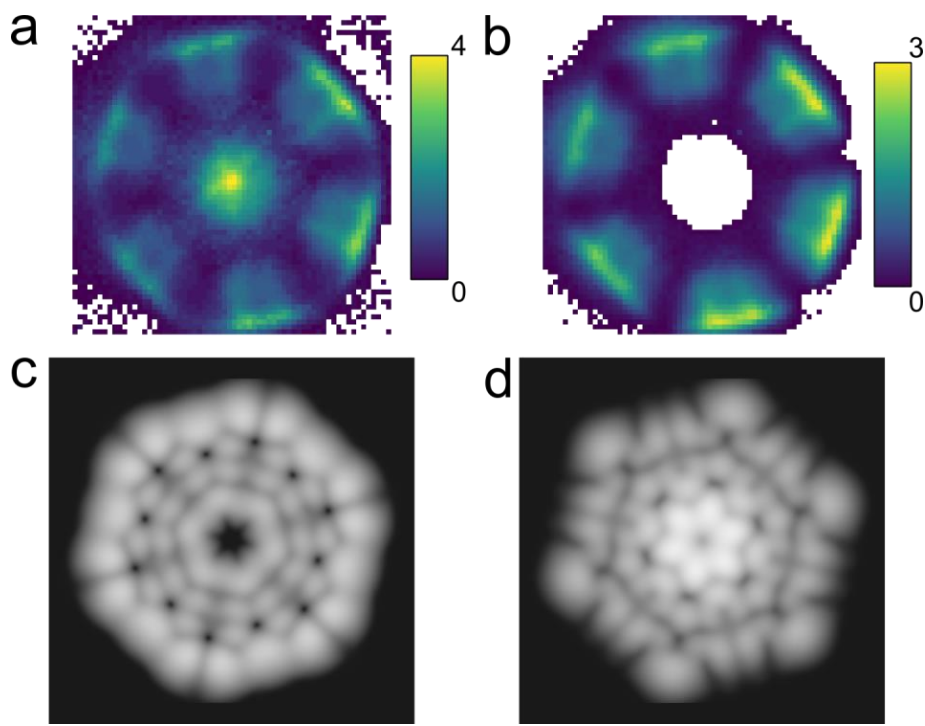

**Supplementary Figure 7. Maps resulting from peak fitting for NIR and calculated molecular orbitals. (a,b)** Feature distribution maps obtained after fitting the $NIR_1$, located at 1.8V, with two Lorentzians whose center have been fixed to 1.84 V and 1.98 V. **(c,d)** Visualization of the molecular wavefunctions as calculated by DFT using an sp-wave mixed tip at the resonance energies $V_{NIR1}$ = 1.82 V and 1.91 V (isovalue 0.0001 a.u.$^2$). Images sizes are 2.4 × 2.4)nm$^2$.

**Synthetic details of the precursor for C108.**

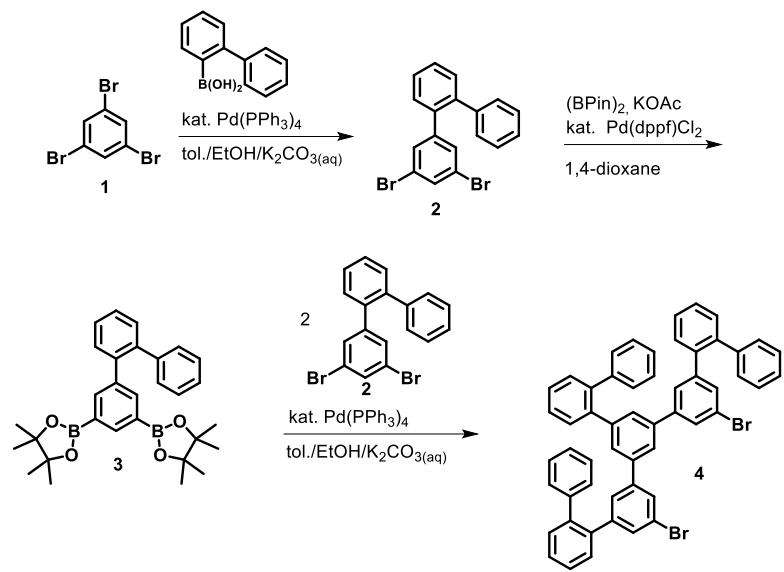

**Scheme 1.** Synthetical route to cyclic nanographene precursor **4**, 5'''-([1,1'-biphenyl]-2-yl)-5'',5''''-dibromo-1,1':2',1'':3'',1''':3''',1'''':3'''',1''''':2''''',1''''''-septiphenyl.

Scheme 1 shows the procedure for the formation of Precursor **4**. In the first step of the three-step synthesis, 1,3,5-tribromobenzene (**1**) was reacted with one equivalent of [1,1'-biphenyl]-2-ylboronic acid to form terphenyl **2** under Suzuki-Miyaura cross coupling conditions. The corresponding boronic acid pinacol ester **3** was synthesized form **2** by Miyaura borylation reaction. The final reaction step for the formation of precursor **4** was accomplished by reacting **3** and **2** in 2:1 ratio under Suzuki Miyaura cross coupling conditions.

**General Information**

All reactions were carried out under inert atmosphere (nitrogen) using Schlenk techniques if not mentioned otherwise. All reagents were purchased from commercial sources if not mentioned otherwise and were used without further purification. For thin-layer chromatography, TLC plates from Merck KGaA with silica gel 60 on aluminum with fluorescence-quenching F254 at room temperature were used. All solvents were dried and/or purified according to standard procedures and stored over 3 Å or 4 Å molecular sieves. NMR spectra were recorded in automation or by the service department (faculty of Chemistry, Philipps University Marburg) with a Bruker Avance 300 or 500 spectrometer at 298 K using $CD_2Cl_2$ or $CDCl_3$ as solvent and for calibration (residual proton signals). HR-APCI mass spectra were acquired with a LTQ-FT Ultra mass spectrometer (Thermo Fischer Scientific). The resolution was set to 100.000. HR-EI mass spectra were ac- quired with a AccuTOF GCv 4G (JEOL) Time of Flight (TOF) mass spectrometer. An internal or external standard was used for drift time correction. The LIFDI ion source and FD-emitters were purchased from Linden ChroMasSpec GmbH (Bremen, Germany). IR spectra are recorded with a Bruker Alpha FT-IR spectrometer with Platinum ATR sampling.

## Chemical synthesis of Precursor 5 and its macrocyclization

3,5-dibromo-1,1':2',1''-terphenyl (**2**)

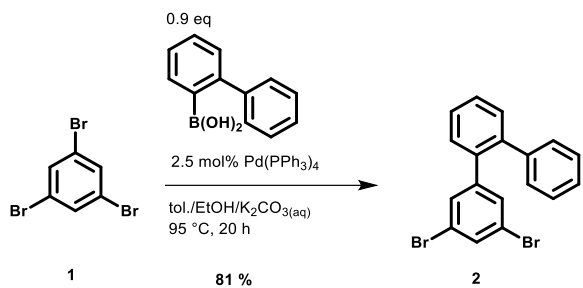

3.42 g (11.0 mmol, 1.00 eq) 1,3,5-tribromobenzene, 2.02 g (10.0 mmol, 0.90 eq) [1,1'-biphenyl]-2-ylboronic acid, 4.15 g (30.0 mmol, 3.00 eq) potassium carbonate and 300 mg (0.25 mmol, 2.5 mol%) tetrakis triphenylphosphino palladium (0) were dissolved in 70 mL toluene, 20 mL ethanol and 10 mL water and degassed. The reaction solution was stirred at 95 °C for 20 h and after this complete conversion could be detected by TLC. The organic phase was separated after cooling to room temperature and the aqueous phase was extracted twice with ethyl acetate (20 mL). The combined organic phases were dried over magnesium sulphate. The crude product was purified by column chromatography on silica gel (eluent: *n*-hexane). 3.13 g (8.11 mmol, 81%) of the desired compound **2** were obtained as colorless solid.

**TLC:** $R_f$ = 0.45 (*n*-hexane)

**¹H NMR:** 300 MHz, 298 K, CDCl$_3$, $\delta$ = 7.53 (t, 1H, $^4J$ = 1.66 Hz), 7.49-7.39 (m, 4H), 7.38-7.29 (m, 3H), 7.24 (d, 2H, $^4J$ = 1.66 Hz), 7.17-7.14 (m, 2H) ppm.

**¹³C NMR:** 75 MHz, 298 K, CDCl$_3$, $\delta$ = 145.6, 141.1, 140.9, 137.9, 132.3, 132.1, 131.1, 130.5, 130.2, 128.9, 128.5, 128.1, 127.3, 122.5 ppm.

**HRMS:** *m/z* for [C$_{18}$H$_{12}$$^{81}$Br $^{79}$Br]$^+$ calc.: 387.92853, found: 387.93066 (EI+).

**IR (ATR):** $\tilde{v}$ = 3061 (w), 3018 (m), 2955 (w), 2896 (w), 1960 (w), 1890 (w), 1591 (w), 1485 (w), 1591 (m), 1485 (w), 1465 (w), 1377 (w), 1250 (s), 1192 (m), 1112 (m), 1024 (w), 1003 (w), 844 (vs), 796 (m), 725 (m), 693 (w), 578 (w), 512 (w) cm$^{-1}$.

**m. p.:** 104-107 °C

3,5-bis(4,4,5,5-tetramethyl-1,3,2-dioxaborolan-2-yl)-1,1':2',1''-terphenyl (**3**)



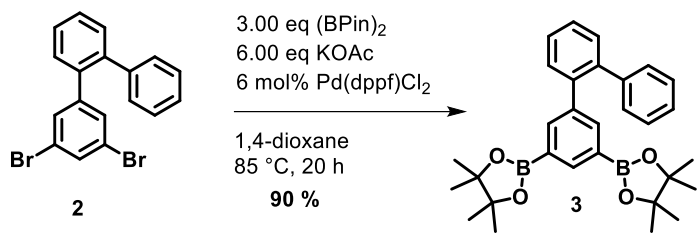

1.50 g (3.89 mmol, 1.00 eq) **2**, 2.98 g (11.67 mmol, 3.00 eq) bis(pinacolato)diboron, 2.30 g (23.34 mmol, 6.00 eq) water-free potassium acetate and 200 mg (0.26 mmol, 6 mol%) 1,1'-bis(diphenylphosphino)ferrocenedichloropalladium(II) were dissolved in 20 mL 1,4-dioxane and stirred at 85_C for 20 h and after this complete conversion could be detected by TLC. After cooling to room temperature, 50 mL water and 50 mL ethyl acetate were added and the aqueous phase was separated and extracted twice with ethyl acetate (20 mL). The combined organic layers were dried over magnesium sulphate and filtered. The crude product was purified by column chromatography on silica gel (eluent: *n*-hexane/ethyl acetate 5:1). 1.69 g (3.50 mmol, 90%) of the desired compound **3** were obtained as colorless solid.

**TLC:** $R_f$ = 0.50 (*n*-hexane/ethyl acetate 5:1)

**$^1$H-NMR:** 300 MHz, 298 K, CD$_2$Cl$_2$, $\delta$ = 7.97 (t, 2H, $^4J$ = 1.09 Hz, *H*4), 7.62 (d, 2H, $^4J$ = 1.09 Hz, *H*2, *H*6), 7.45-7.39 (m, 4H), 7.23-7.13 (m, 5H), 1.29 (s, 24H, *H*3', *H*5') ppm.

**$^{13}$C-NMR:** 75 MHz, 298 K, CD$_2$Cl$_2$, $\delta$ = 144.2, 142.0, 141.1, 140.9, 140.6, 139.6, 139.2, 130.8, 130.7, 130.3, 128.1, 127.7, 126.8, 84.1, 25.0 ppm

**HRMS:** *m/z* for [C$_{20}$H$_{36}$$^{11}$B$_2$$^{16}$O$_4$]$^+$ calc.: 482.27997, found: 482.28049 (EI+).

**IR (ATR):** $\tilde{v}$ = 2977 (m), 2928 (w), 2871 (w), 1699 (w), 1582 (w), 1557 (w), 1478 (w), 1441 (w), 1407 (w), 1370 (s), 1347 (vs), 1317 (s), 1269 (s), 1213 (w), 1112 (m), 1069 (vs), 1008 (m), 964 (m), 867 (m), 845 (w), 776 (m), 700 (s), 676 (w), 616 (w), 578 (w) 517 (w) cm$^{-1}$.

**m. p.:** 105-108 °C

5'''-([1,1'-biphenyl]-2-yl)-5'',5''''-dibromo-1,1':2',1'':3'',1''':3''',1'''':3'''',1''''':2''''',1'''''''-septiphenyl (**4**)

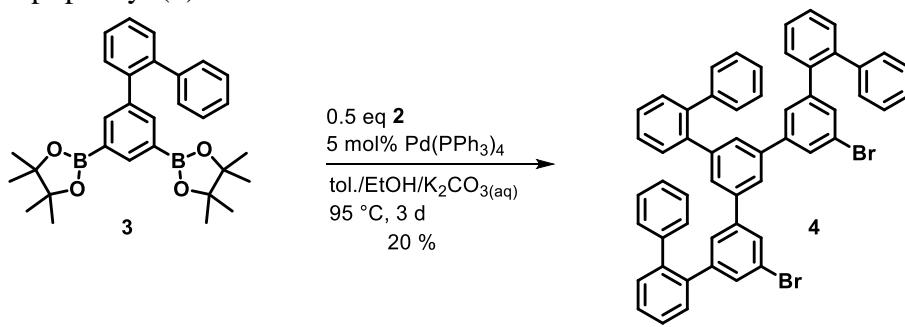



794 mg (1.65 mmol, 0.50 eq) (3,5-dibromo-1,1':2',1''-terphenyl (**2**), 1.27 g (3.30 mmol, 1.00 eq) boronic acid pinacol ester **3**, 1.37 g (9.90 mmol, 6.00 eq) potassium carbonate and 95 mg (0.08 mmol, 6 mol%) tetrakis triphenylphosphino palladium (0) were dissolved in 50 mL toluene, 20 mL ethanol and 10 mL water and degassed. The reaction solution was stirred at 70_C for 3 d and after this complete conversion could be detected by TLC. The organic phase was separated after cooling to room temperature and the aqueous phase was extracted twice with dichloromethane (20 mL). The combined organic phases were dried over magnesium sulphate. The crude product was purified for three times by column chromatography on silica gel (eluent: *n*-pentane/DCM 4:1). 280 mg (8.33 mmol, 20%) of the desired compound **4** were obtained as colorless solid.

**TLC:** $R_f$ = 0.30 (*n*-pentane/dichloromethane 10:1)

**$^1$H - NMR:** 300 MHz, 298 K, CD$_2$Cl$_2$, $\delta$ = 7.55-6.97 (m, 36H) ppm.

**$^{13}$C - NMR:** 75 MHz, 298 K, CD$_2$Cl$_2$, $\delta$ = 144.2, 143.9, 143.5, 142.6, 142.4, 142.1, 141.9, 141.6, 141.5, 141.2, 140.2, 139.7, 139.6, 134.5, 131.8, 131.0, 130.9, 130.7, 130.7, 130.4, 130.4, 129.0, 128.8, 128.7, 128.7, 128.5, 128.5, 128.3, 128.1, 128.0, 128.0, 127.3, 125.7, 124.0, 122.7 ppm.

**HRMS:** *m/z* for [C$_{54}$H$_{36}$$^{81}$Br$^{79}$Br]$^+$ calc.: 844.11633, found: 844.11443 (EI+).

**IR (ATR):** $\tilde{v}$ = 3057 (w), 2925 (w), 2854 (w), 1733 (w), 1591 (m), 1564 (m), 1476 (m), 1444 (w), 1389 (m), 1295 (w), 1262 (w), 1159 (m), 1110 (w), 1071 (w), 1015 (w), 865 (m), 805 (w), 738 (vs), 700 (vs), 659 (m), 617 (m), 519 (w) cm$^{-1}$.

**m. p.:** 120-123 °C

**(b) NMR Spectra**

(3,5-dibromo-1,1':2',1''-terphenyl (**2**))

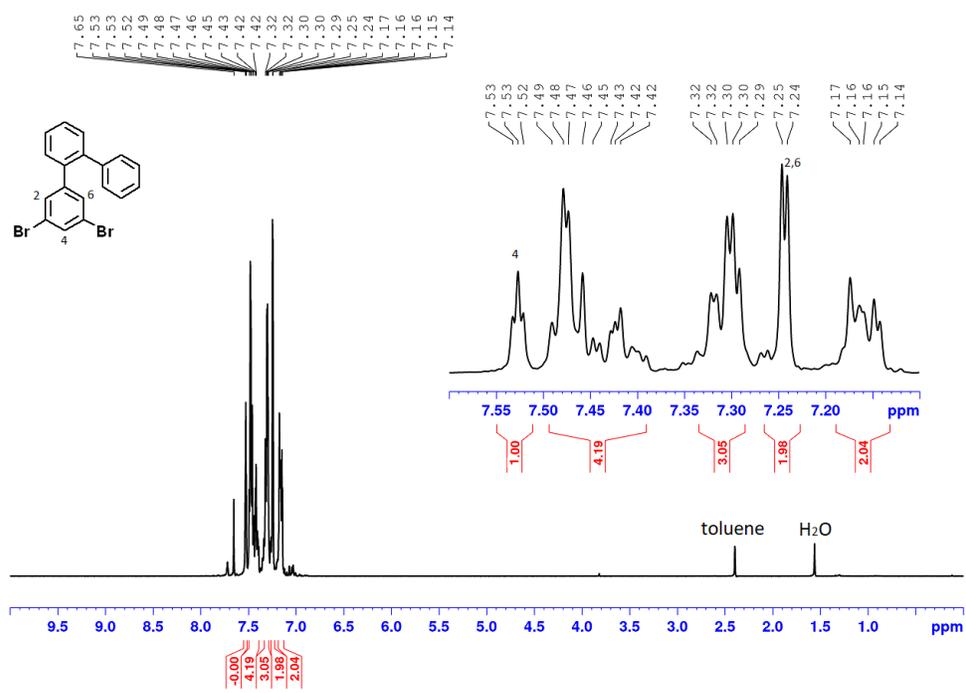

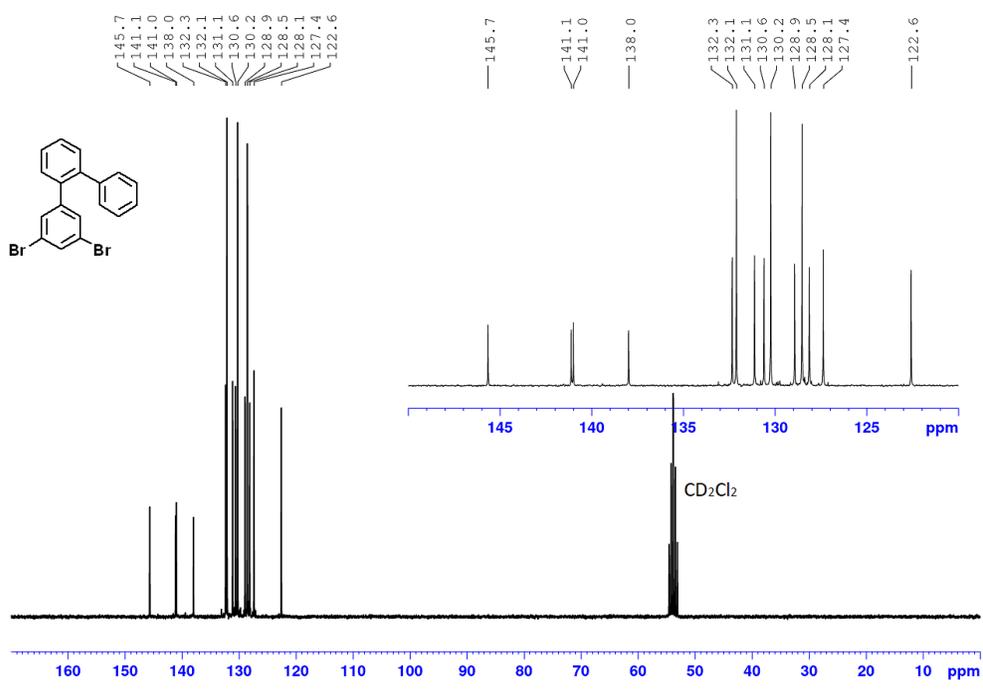

3,5-bis(4,4,5,5-tetramethyl-1,3,2-dioxaborolan-2-yl)-1,1':2',1''-terphenyl (**3**)

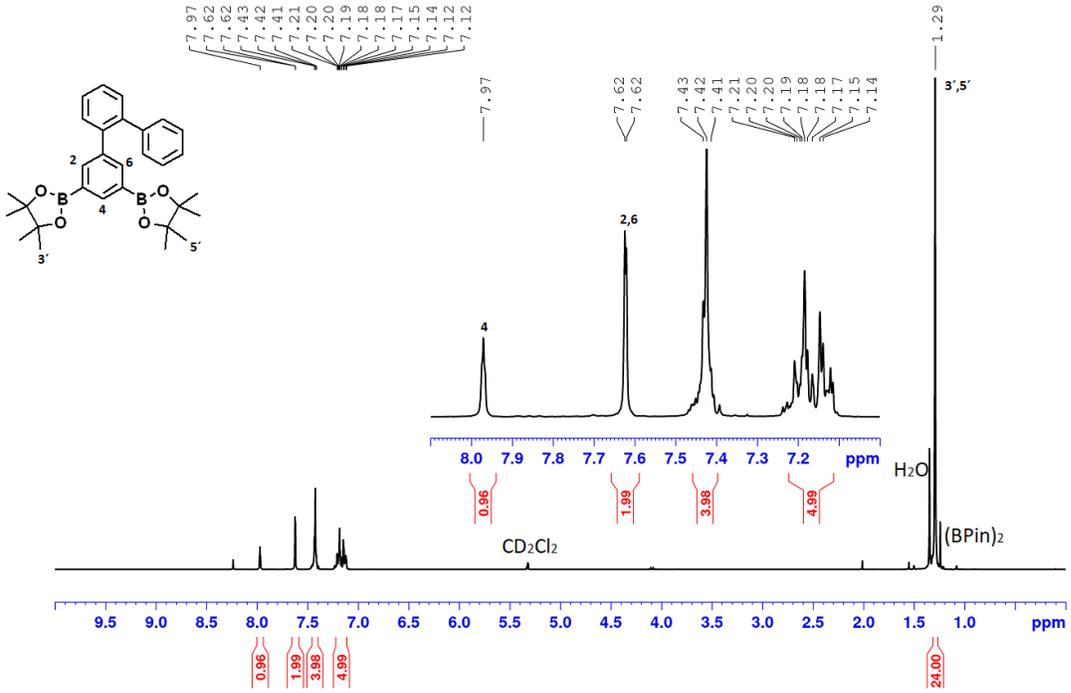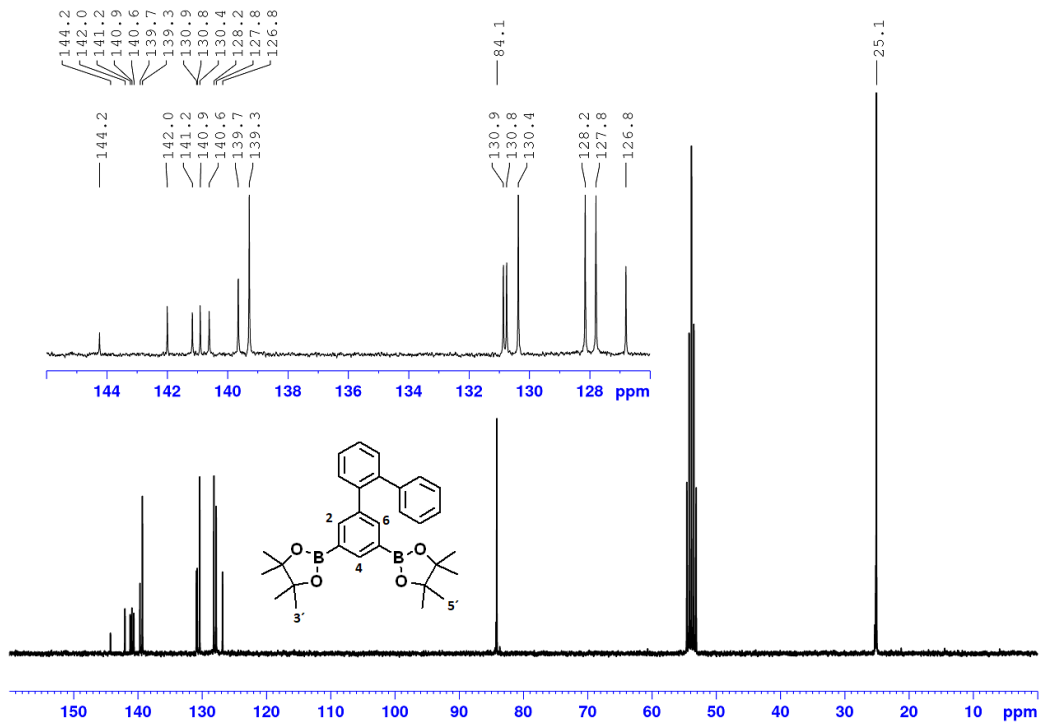

5'''-([1,1'-biphenyl]-2-yl)-5'',5''''-dibromo-1,1':2',1'':3'',1''':3''',1'''':3'''',1''''':2''''',1'''''''-sepiphenyl (**4**)